\documentstyle[12pt,epsf]{article}
\def\thebibliographyss#1{\subsubsection*{References}\list
  {[\arabic{enumi}]}{\settowidth\labelwidth{[#1]}\leftmargin\labelwidth
    \advance\leftmargin\labelsep
    \usecounter{enumi}}
    \def\newblock{\hskip .11em plus .33em minus -.07em}
    \sloppy
    \sfcode`\.=1000\relax}

\begin{document}

\vspace*{-2 in}
\begin{flushright}
MC-TH-96/03\\
LU TP 96-4\\
UCL/HEP 96-01
\end{flushright}

\begin{center}
{\bf Multiple Hard Parton Interactions at HERA}
\bigskip

{\bf J M Butterworth$^1$, J R Forshaw$^2$, T Sj\"ostrand$^3$, and 
J K Storrow$^2$ }

\medskip
$^1${\sl Department of Physics and Astronomy, University College London,
London, UK}

$^3${\sl Department of Theoretical Physics, University of Manchester,
Manchester, UK}

$^2${\sl Department of Theoretical Physics, University of Lund,
Lund, Sweden }

\end{center}

{\bf Abstract}

At HERA, the large flux of almost real photons accompanying the
electron beam leads to the copious photoproduction of jets.
Regions of small momentum fractions $x$ of the incoming particles are 
explored, where the density of partons is high.
As a result, the probability for more than one hard partonic
scattering occurring in a single $\gamma p$ collision could become
significant. It is well known that this effect modifies the
contribution of jets (minijets) to the total cross section.
We discuss the latest HERA data on the total $\gamma p$ cross section
in this context.
The possible effects of multiple hard interactions on event shapes 
and jet cross sections at HERA have been studied using Monte Carlo  programs. 
We review some of the available results, which in general indicate that 
the effects of multiple interactions should be significant and may 
already be manifest in the existing HERA data.

\normalsize\baselineskip=15pt
\setcounter{footnote}{0}
\renewcommand{\thefootnote}{\alph{footnote}}

{\bf 1. Introduction}

For both protons and photons, QCD predicts a rapid increase in parton 
densities at low $x$, where $x$ is the fraction of the beam 
particle's momentum which participates in the `hard' scattering (interaction).
In a naive treatment, this rise can lead to a corresponding (but
ultimately unphysical) rise with increasing energy of perturbative QCD
calculations of the jet contribution to the total cross section.
However, the large number of small $x$ partons
contributing to jet production can mean that there is a significant
probability for more than one hard scatter per $\gamma p$
interaction. Only the resolved part of the photon can undergo 
multiple interactions, i.e. the direct part is taken to be unaffected.
The effects of multiple interactions can provide a mechanism for 
taming the rise in the QCD cross section \cite{guys} in accord
with unitarity, as discussed in Sj\"ostrand's talk \cite{Torbjorn}.

\begin{figure}
\begin{center}
\leavevmode
\hbox{\epsfysize=1.75 in
\epsfxsize= 2.5 in
\epsfbox{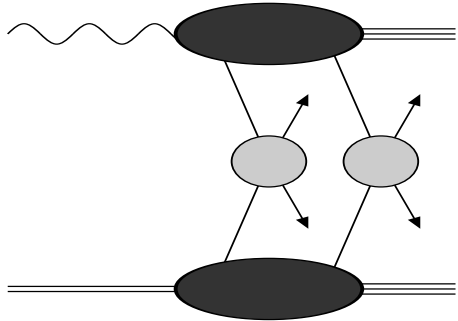}}
\end{center}
\vspace{-5mm}
\caption[]{An example of a multiple scattering in a $\gamma p$ collision.}
\end{figure}

Additionally, there are clearly implications for the hadronic final state.
In order to study these effects eikonal models
have been implemented within the hard process generation of several
Monte Carlo programs~\cite{HERWIG,honest,PYTHIA,PHOJET}.
Multiple parton scattering, as illustrated in Fig.1, is expected to
affect jet rates.
The average number of jets per event should be increased when partons
from secondary hard scatters are of sufficiently high $p_T$ to give jets
in their own right. In addition, lower $p_T$ secondary scatters
produce extra transverse energy in the event which contributes to
the pedestal energy underneath other jets in the event.
Thus multiple scattering can influence jet cross sections even when
no parton from the secondary scatters is itself of a high enough $p_T$
to produce an observable jet. By boosting the transverse energy of jets
in this way, multiple scattering leads to an increase in jet cross sections
for jets above a certain $E_T^{jet}$ cut, even though the total
cross section is reduced.

The theory of multiple scattering is still not well understood; therefore
phenomenology is based on models with several assumptions and unknown
parameters. The effects of multiple scattering on the total cross section
and on jet/event shapes probe these unknowns in somewhat different ways.
Although they are related, it therefore makes sense to consider the two 
aspects separately. In sect.2 we outline the framework of models of 
multiple interactions, in sect.3 we discuss the HERA total cross section 
data for $\gamma p$
quantitatively from the minijet point of view. In sect.4
we present some of the expected effects of multiple interactions on jet
cross sections, with reference to existing data, and in sect.5 we give
some conclusions.

{\bf 2. Modelling multiple interactions}

When speaking of multiple scattering (alias multiple interactions),
the basic building blocks are the standard $2 \to 2$ partonic 
interaction processes, such as  $g g \to gg$, $q q \to q q$, $q \overline{q} 
\to g g$ and $q g \to q g$. An event may contain none, one, two or more such
interactions; the novelty of the multiple interactions concept
is that the `two or more' occurrence may be the norm at high energies
rather than a rare exception. The correlations between interactions
occurring in the same event may be horribly complicated, but the hope is 
that it is feasible to approximate this by a simple factorized description 
in terms
of several independent scatterings. This may be made plausible by viewing
the incoming hadron and hadronlike photon as two pancakes, flattened
by Lorentz contraction. When these pass through each other, different
parts are causally separated, so the probability of an interaction 
between any pair of partons can be assumed independent of what
occurs anywhere else. This gives a poissonian distribution in the
number of interactions, for a fixed impact parameter between the two
colliding hadrons/photons. In central collisions the average number 
of interactions should be larger than in peripheral ones; it is therefore
necessary to introduce a model of how partons are distributed 
in the pancake. The `eikonalization' procedure, so central to
current descriptions, is a combination of these two concepts:
a poissonian distribution at each fixed impact parameter plus an
impact-parameter-dependent overlap function. This form may be integrated
over the impact parameter to give the total cross section, the probability
distribution of interactions, and even the number of additional
interactions underlying a hard interaction. 

Several objections can be raised. For instance, scatterings are assumed
to be disjoint, i.e. a parton does not undergo more than one scattering.
This can be motivated by simple counting arguments: if each hadron
contains $N$ partons, the rate of two disjoint $2 \to 2$ scatterings
is proportional to $N^4$ but that of two scatterings with a shared parton, 
i.e. a $3 \to 3$ process, is only proportional to $N^3$. A counterexample
would be the occurrence of `hot spots' caused by the cascading of a
single parton. This is especially relevant since each $2 \to 2$ 
subprocess is embedded in a larger system of initial- and 
final-state radiation. Colour correlations between interactions are not
well understood, and yet this is of large importance for the structure
of the final event. In summary, current eikonalization approaches can  
only be viewed as sensible first approximations, and therefore the
existence of several different models is an advantage.

Maybe the most ambitious approach is the DTU (dual topological 
unitarization) one. Here a hard pomeron term is given by
scatterings with $p_T > p_{Tmin}$, a soft pomeron term includes 
nonperturbative scatterings at small $p_T$, and triple- and 
loop-pomeron graphs are added to incorporate also diffractive 
topologies in the description. The $p_{Tmin}$ scale can be changed
freely over some range, since the nonperturbative pomeron term can 
be modelled to compensate for the change in the hard interaction rate. 
This approach is represented by the {\sc Phojet} program \cite{PHOJET}.

In less ambitious approaches, the issue of diffractive events is kept
separate, and unitarization within the soft sector is simplified to
allow (at most) one soft interaction. Here $p_{Tmin}$ is a real
parameter of the theory, to be determined by a comparison with data.
Of course the use of a sharp cutoff in $p_T$ is only an approximation
to what has to be a smooth turn on in reality.
Examples include the models found in {\sc Pythia} \cite{PYTHIA} and
{\small HERWIG} \cite{HERWIG,honest}. These two are similar in philosophy
but differ in the details. For simplicity we therefore concentrate on
the latter in the following.

Let us start by considering a $\gamma p$ interaction at some fixed
impact parameter, $b$, and centre-of-mass energy, $s$. In particular, 
we suppose the $\gamma$ to be hadronlike (i.e. resolved \cite{Kramer}). 
We assume that the probability for the
photon to interact in such a state is $P_{res}$ and take the $\rho$-dominance
form, i.e. $P_{res} = 4 \pi \alpha_{{\rm em}}/f_{\rho}^2 \approx 1/300$. 
The mean number of jet pairs produced in this resolved-$\gamma$--$p$ 
interaction is then 
\begin{equation}
\langle n(b,s) \rangle = {\cal L}_{{\rm partons}} \otimes \hat{\sigma}_H
\end{equation}
where ${\cal L}_{{\rm partons}}$ is the parton luminosity and $ \hat{\sigma}_H$
is the cross section for a pair of partons to produce a pair of jets (i.e.
partons with $p_T > p_{Tmin}$).

The convolution is because the parton cross section depends upon the parton
energies. More specifically,
\begin{equation}
d {\cal L}_{{\rm partons}} = A(b) 
n_{\gamma}(x_{\gamma}) n_p(x_p) dx_{\gamma} dx_p
\end{equation}
where $A(b)$ is a function which specifies the
distribution of partons in impact parameter. It must satisfy
$$ \int \pi db^2 A(b) = 1$$ in order that the parton luminosity integrated
over all space is simply the product of the parton number densities.
Factorizing the $b$ dependence like this is an assumption. In particular
we do not contemplate QCD effects which would spoil this, e.g. perhaps
leading to \lq hot spots' of partons.
Also $n_i(x_i)$ is the number density of partons in hadron $i$ which carry
a fraction $x_i$ of the hadron energy. For ease of notation we do not
distinguish between parton types and have ignored any scale dependence of
the number densities. For the proton, the number density is none other than
the proton parton density, i.e. $n_p(x_p) \equiv f_p(x_p)$. However, since
we are dealing with resolved photons, the number density $n_{\gamma}$ is
related to the photon parton density by a factor of $P_{res}$, i.e.
$n_{\gamma}(x_{\gamma}) = f_{\gamma}(x_{\gamma})/P_{res}$.
Thus, after performing the convolution, we can write:
\begin{equation}
\langle n(b,s) \rangle = \frac{A(b)}{P_{res}} \sigma_H^{inc}(s),
\end{equation}
where $\sigma_H^{inc}(s)$ is the inclusive cross section for $\gamma p$ to
jets. Restoring the parton indices, it is given by
\begin{equation}
\sigma^{inc}_H(s) = \int_{p_{Tmin}^2}^{s/4}
\hspace{-0.4cm} dp_T^2 \int_{4p_T^2 /s}^{1} \hspace{-0.4cm}
 dx_\gamma \int_{4p_T^2 / x_{\gamma} s}^{1} \hspace{-0.5cm} dx_p \sum_{ij}
f_{i/\gamma }(x_\gamma,p_T^2) f_{j/p}(x_p,p_T^2) \;
\frac{d\hat{\sigma}_{ij}(x_\gamma x_p s,p_T)}{dp_T^2}.
\end{equation}

In order to investigate further the structure of events containing 
multiple interactions we need to know the probability distribution for
having $m$ (and only $m$) scatters in a given resolved-$\gamma$--$p$ event,
$P_m$. In order to do this we assume that the separate scatters are
uncorrelated, i.e. they obey poissonian statistics. 
Thus
\begin{equation}
P_m = \frac{(\langle n(b,s) \rangle)^m}{m!} \exp (-\langle n(b,s)\rangle).
\end{equation}
This formula is central to the Monte Carlo implementation in {\small HERWIG}.

We can now ask for the total cross section for $\gamma p \to$ partons with
$p_T > p_{Tmin}$.
\begin{eqnarray}
\sigma_H(s) &=& \pi P_{res} \int db^2 \sum_{m=1}^{\infty} P_m \nonumber \\
&=& \pi P_{res} \int db^2 [1 - \exp(-\langle n(b,s)\rangle)].
\end{eqnarray}

Since the total inclusive cross section ($\sigma_H^{inc}$) counts all jet 
pairs (even ones which occur in the same event) we expect it to be larger than
$\sigma_H$ by a factor equal to the mean number of multiple interactions
per event (i.e. averaged over impact parameter). This is easy to see. Let
$\langle n(s) \rangle$ be the average number of jet pairs produced in
resolved-$\gamma$--$p$ events which contain at least one pair of jets, then
\begin{eqnarray}
\langle n(s) \rangle &=& \frac{ \int db^2 \sum_{m=1}^{\infty} m P_m }{
\int db^2 \sum_{m=1}^{\infty} P_m} \nonumber \\
&=& \frac{ \int db^2 \langle n(b,s) \rangle }{
\int db^2 [1 - \exp( -\langle n(b,s)\rangle)]} \nonumber \\
&=& \frac{ \sigma_H^{incl}(s)}{\sigma_H(s)}.
\end{eqnarray}
Note that $\sigma_H$ must always be less than the total $\gamma p$ cross
section, whereas $\sigma_H^{incl}$ need not be. 

In order to study the details of the hadronic final state in the presence of
multiple interactions it is most convenient to use a Monte Carlo simulation.
The {\small HERWIG} Monte Carlo generates the required number of
hard scatters and the associated initial and final state parton showering.
The outgoing partons and remnant jets are then fragmented to
the hadronic final state.
Two modifications are made to the simple eikonal model in the implementation.
Firstly, energy conservation is imposed, i.e.
after the backward evolution of all the hard scatters in
an event, the energy remaining in the hadronic remnants
must be greater than zero. Secondly, if during the backward evolution of the
first scatter the splitting $q\bar{q} \leftarrow \gamma$ is arrived
at before the evolution cut off scale, the event is classified as an
`anomalous' event and no multiple interactions are allowed.

{\bf 3. Total $\gamma p$ cross section}

The `minijet' contribution, $\sigma_H(s)$, is clearly only one contribution to
the total cross section. To compare with experiment we must add the soft 
cross section $\sigma_T^{soft}$, which is assumed to contain all of the 
physics below $p_{Tmin}$. Furthermore there is a (small) cross 
section, $\sigma_T^{dir}$, for the production of 
single jet pairs (with $p_T > p'_{Tmin}$) by unresolved (i.e. direct) 
photons.  Often it is assumed that $p'_{Tmin} = p_{Tmin}$, but this differs 
between models. Of course simply adding together cross sections is only a
first guess.
 
As an example of multiple interaction approaches to the total
$\gamma p$ cross section, we take the eikonalized minijet calculations for
$\sigma_H(s)$ of ref.\cite{FS1}, which used the parton distribution functions
of refs.\cite{MT,DG} for the proton and photon respectively, and
add to them various choices of parametrisations of $\sigma_T^{soft}$. We
adjust $p_{Tmin}$ and the parameters of $\sigma_T^{soft}$ accordingly, and
compare to the recent data on $\sigma_T^{\gamma p}$ \cite{H1N,ZEUSN}.
This is similar to what was done in ref.\cite{FS2}, except there the old
HERA data were used \cite{H1,ZEUS}.
For a fuller discussion of this approach and its theoretical
uncertainties see refs.\cite{Torbjorn,FS1,FS2}.

The recent results for the total $\gamma p$ cross
section at centre of mass energies of $\simeq 200$ GeV are as follows. The H1
collaboration
$(\langle W \rangle =\langle \surd s \rangle = 197$ GeV) find
$\sigma_T^{\gamma p} = (165 \pm 2.3 \pm 10.9)\, {\rm \mu b}  $
and ZEUS $(\langle W \rangle = 180$ GeV) obtain
$\sigma_T^{\gamma p} = (143 \pm 17)\, {\rm \mu  b}  $.
In the H1 case, the first error is statistical and the second systematic. The
ZEUS error has the systematic and statistical errors added in quadrature.

\begin{figure}[t]
\begin{center}
\leavevmode
\hbox{\epsfysize=4.0 in
\epsfbox{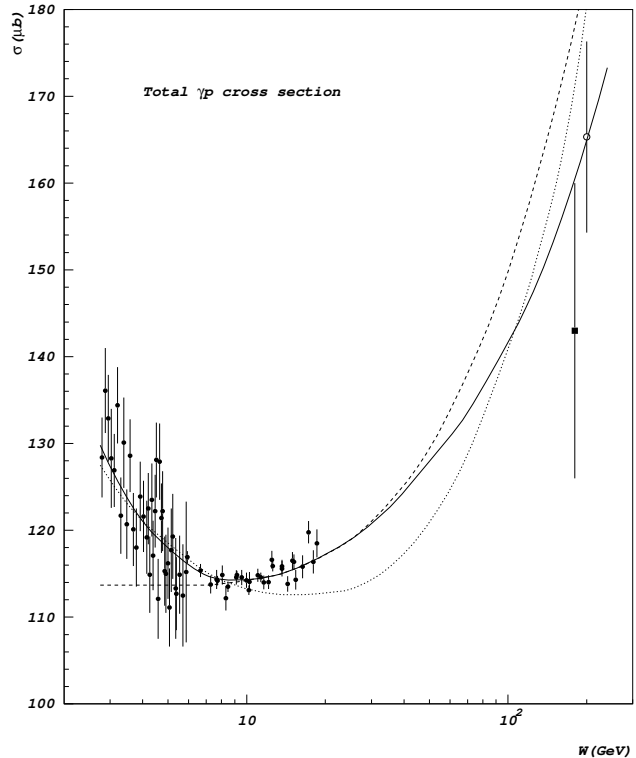}}
\end{center}
\vspace{-5mm}
\caption[]{Energy dependence of the total $\gamma p$ cross section.
See text for the description of the curves.}
\end{figure}

In the simplest variant, $\sigma_T^{soft}$ is taken to have the form:
\begin{equation}
\sigma_T^{soft}(s) = A  + B/ \surd s
\end{equation}
where $A$ and $B$ are constants. This ansatz implies that the entire rise in
$\sigma_T^{\gamma p}$ at high energies is due to minijets. It has very
little theoretical motivation and makes rather an ad hoc separation between
hard and soft physics, endowing $p_{Tmin}$ with great physical significance.
In fig.2 we show as the dotted curve the best fit to the data
that can be achieved with such an ansatz added to the results of
ref.\cite{FS1} for $\sigma_H(s)$, calculated with $p_{Tmin} = 2$ GeV .
We take $A = 107.98\, {\rm \mu}$b and $B = 54.34 \,{\rm \mu}$b GeV. This
leads to an unconvincing description of the
the data in the 10--18 GeV range \cite{data}, which show
evidence of a rise with energy. If one attempts to attribute the low energy
rise entirely to the minijet contribution then one would be forced to choose
very low values of $p_{Tmin}$, in the 1--1.5 GeV range (see 
ref.\cite{Torbjorn}), which would lead to
much too high a prediction at HERA, as was found in ref.\cite{Honjo}. This is
illustrated by the broken curve in fig.2, where we take a constant
$\sigma _T^{soft}=114 {\rm \mu}$b and
$p_{Tmin} = 1.5$~GeV in $\sigma_H(s)$. Again the need to fit
the low energy rise leads to a dangerously high cross section at HERA
energies (no attempt was made to fit the data below $\simeq 10$ GeV: any 
attempt to do so must worsen the fit to the higher energy data).

Also in fig.2 we show (the solid line) the results of ref.\cite{FS1} added 
to the following form for $\sigma_T^{soft}$
\begin{equation}
\sigma_T^{soft}(s) = A s^\epsilon + B/\surd s .
\end{equation}
The choice $A = 78.4\, {\rm \mu}$b, $\epsilon =0.058\,$
and $B = 117.05 \,{\rm \mu}$b GeV, with $p_{Tmin} = 3$ GeV in
$\sigma_H (s)$, ensures an excellent fit to the low energy
data (i.e. $W\le 18$ GeV) \cite{data}.

This latter approach clearly provides the best description of the data, but is
slightly unorthodox in attributing only part (around half) of the rise in
cross section to minijets. We could argue that this is a rational view of 
the situation, less extreme than attributing all
of the rise to $\sigma_T^{soft}$, as in the soft Pomeron 
approach \cite{DL}. We emphasize that the minijet question is
not a stark choice between the two extremes of either minijets providing
the entire rise with energy of the total cross section or being absent:
indeed, jets are undeniably produced in both $ {\overline {p}}p $
\cite{UA} and $\gamma \gamma$ \cite{GG}, as well as $\gamma p$, reactions.
The important issue is whether they make a significant contribution to the
total cross section at existing energies. In this picture, with 
$p_{Tmin} = 3\,$GeV, they do; $\sigma_H (s)$  
being around 20 
${\rm \mu}$b at $\surd s =200 $ GeV,
although this is very sensitive to the choice of $p_{Tmin}$. 
The increase of the soft cross section could still have its origin
in multiple interactions in the region $p_T < p_{Tmin}$,
but then without a simple perturbative description.

{\bf 4. Hadronic final state}

In this section we illustrate the possible size and nature of the effect
of multiple interactions on some aspects of jet cross sections and the
hadronic final state at HERA.
Here the experiments typically measure jets with transverse energies as
low as 6~GeV. To simulate these events with Monte Carlo models, they use
values for the minimum transverse momentum of a hard scatter of around
$2 < p_{Tmin} <  3$ GeV. In addition, cuts on the $\gamma p$ CM energy are
made, forcing it to lie typically in a range of around
120 GeV $\leq \sqrt{s} \leq$  265 GeV. For these choices the mean
number of hard scatters per event was found to be at least 1.04 
(higher for the lower
$p_{Tmin}$ values), taking the MRS $D_{-}$~\cite{MRS} proton and 
GRV~\cite{GRV} photon parton densities.

For the events we generate, since the cross section falls rapidly with
increasing $p_T$, we expect that most of the multiple interactions will have
$p_T \sim p_{Tmin}$ and so their main effect will be to increase the mean
number of observed jets by boosting the underlying $E_T$ in
the event. Events which contain multiple interactions with high enough $p_T$
to be observed as jets in their own right are relatively rare, but their
observation (they appear as pairs of back-to-back jets) would provide
striking evidence in support of the existence of multiple interactions.

\begin{figure}
\begin{center}
\leavevmode
\hbox{\epsfxsize=15 cm
\epsfbox{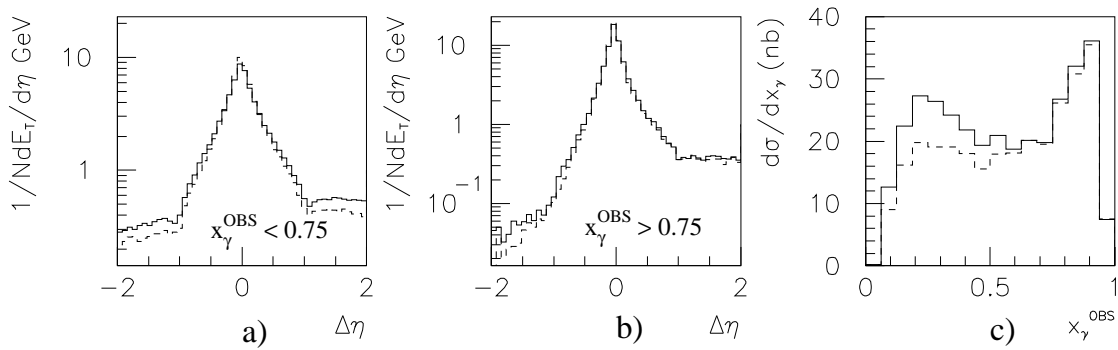}}
\end{center}
\vspace{-5 mm}
\caption[]{
The $E_T$ jet profile in $\eta$,
for a) $x_\gamma^{obs}< 0.75$ and b) $x_\gamma^{obs} \ge 0.75$.
c) $x_\gamma^{obs}$ distribution for dijets, at HERA energies,
with direct contribution. Including multiple scattering
(solid line) and with multiple scattering turned off (broken line).
}
\end{figure}

Jet finding was performed using a cone algorithm with a cone radius $R=1$
\cite{cone}. Jets have $E_T \geq 6$ GeV
and pseudorapidity $-2 \leq \eta ^{jet} \leq 2$. The additional $E_T$ can be
seen directly in the jet profile, Fig.3(a), where the $E_T$ in the jets is
 plotted
against $\eta$ relative to the jet axis. The pedestal energy is increased with
the inclusion of multiple interactions. A variable found to be useful at HERA
\cite{xgam} for distinguishing between direct and resolved photon events is 
the `observable $x_\gamma$'. It is defined by
\begin{equation}
x_{\gamma}^{obs} = \frac{\sum_{jets} E_T^{jet}
e^{-\eta^{jet}}}{2E_{\gamma}}
\end{equation}
and the sum is over the two highest $E_T$ jets in the event. It is the 
fraction of the
photon's energy manifest in the two jets of highest $E_T$, and in leading 
order is
exactly the fraction of the photon energy which enters the hard scatter, i.e.
direct events are peaked at $x_{\gamma}^{obs} = 1$ whilst resolved events have
$x_{\gamma}^{obs}< 1$. The $x_\gamma$ cut isolates the resolved interactions,
and as expected multiple interactions have no effect in the direct case, 
Fig.3(b).
The extra $E_T$ enhances the inclusive jet rate around $\eta ^{jet} = 1$, and
an increased sensitivity to multiple scattering can be seen in the 
$\eta ^{jet}$
and $E_T^{jet} $ distributions in dijet events, which are shown in 
ref.\cite{honest}. An
enhancement at high $\eta^{jet}$ and low $E_T^{jet}$ is seen. All plots in 
Fig.3 have
the direct contribution generated using {\small HERWIG} included, hence the 
rise at $x_\gamma^{obs} \sim 0.8$ in Fig.3(c).

It can be seen from Fig.3(c) that the effect of multiple scattering is 
greater in
the low $x_\gamma^{obs}$ region. Multiple interactions are more likely to 
occur here as
it is in this region that the higher parton densities occur; also the energy
conservation constraint is less restrictive.

\begin{figure}[t]
\begin{center}
\leavevmode
\hbox{\epsfysize=200pt
\epsfbox{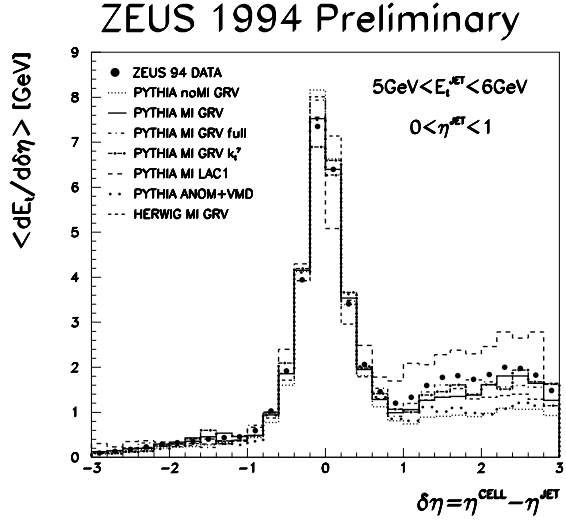}}
\end{center}
\vspace{-5 mm}
\caption[]{
Jet profile. The solid circles are uncorrected ZEUS data.
Uncorrected transverse energy flow seen in the calorimeter
around the jet axis $ \langle dE_T/d\delta\eta \rangle$ where
$\delta\eta = \eta^{cell}-\eta^{jet}$,
for cells within one radian in $\phi$ of the jet axis,
for jets in the range 5~GeV$ < E_T^{jet}< 6$~GeV and with 
$0 < \eta^{jet} < 1$.
}
\end{figure}

In Fig.4 various Monte Carlo models are compared to preliminary ZEUS
data~\cite{ZEUSbrus}.
The jet profile is shown, and the data are not corrected for detector effects.
The simulated events have been passed through a full simulation of the
experiment. Jets have 5~GeV $< E_T^{jet} < 6$~GeV and pseudorapidities in the
range $0 < \eta^{jet} < 1$. These cuts remove the effect of the forward edge 
of
the calorimeter acceptance and allow us to study the effect of the different
models on the jet profile independently of the $E_T^{jet}$ and $\eta^{jet}$
distribution of the models.
The comparison with the various models confirms the general conclusion 
made by H1~\cite{H1jets}
that introducing multiple interactions can improve the agreement between Monte
Carlo models and the data. However, there is a large amount of freedom
in the models and, in particular, the effect of introducing multiple
interactions depends strongly on the parton density of the photon (compare
the histograms for the multiple interaction models using LAC1 and GRV).
For further comparisons with HERA data, the reader is referred
to the talk of Steve Maxfield~\cite{steve} and references therein.

{\bf 5. Conclusions}

Multiple scattering must occur at some CM energy, in order that unitarity is
not violated: the relevant point here is whether it is present at HERA.
Clearly answering this question using total cross section data is going to be
very difficult, given the uncertainties in the soft physics.
This point was made very clear by our discussion of sect.3, and
in ref.\cite{Torbjorn}.

However, jet cross sections seem to be a more hopeful place to look.
In sect.4, the effect on the hadronic final state of multiple
parton scattering in $\gamma p$
interactions has been simulated by interfacing an
eikonal model of multiple parton interactions with {\small HERWIG}.
Models indicate that the effect of
multiple scattering is significant at HERA energies. For
reasonable experimental cuts, the inclusion of multiple
scattering leads to significant changes in inclusive and
dijet cross sections which should be understood before attempting
to unfold to parton distribution functions.

\begin{thebibliographyss}{10}
\bibitem{guys}J. C. Collins and G. A. Ladinsky, Phys.
Rev. {\bf D43} (1991) 2847;
J. R. Forshaw and J. K. Storrow, Phys. Lett. {\bf B268} (1991)
116; erratum {\bf B276}, (1992) 565;
R. S. Fletcher et al., Phys. Rev. {\bf D45} (1992) 337.
\bibitem{Torbjorn}T.~Sj\"ostrand, these proceedings.
\bibitem{HERWIG} G.~Marchesini et al., Comp. Phys. Comm. {\bf 67} 
(1992) 465;
J. M. Butterworth and J. R. Forshaw, J. Phys. {\bf G19} (1993) 1657.
\bibitem{honest}
J. M. Butterworth, J. R. Forshaw and M. H. Seymour, CERN-TH/95-82.
\bibitem{PYTHIA}T. Sj\"ostrand and M. van Zijl, Phys. Rev. {\bf D36} 
(1987) 2019;
G.A. Schuler and T. Sj\"ostrand, Phys. Lett. {\bf B300} (1993) 169,
Nucl. Phys. {\bf B407} (1993) 539;
T. Sj\"ostrand, Comput. Phys. Commun. {\bf 82} (1994) 74.
\bibitem{PHOJET} R. Engel, Z.Phys. {\bf C66} (1995) 203.
\bibitem{Kramer}G. Kramer, these proceedings;
J. K. Storrow, J.Phys. {\bf G19} (1993) 1641.
\bibitem{FS1} J. R. Forshaw and J.K.Storrow, Phys.Rev. {\bf D46}, 4955 (1992).
\bibitem{MT}J. Morfin and W.-K.Tung, Z.Phys. {\bf C52} (1991) 13.
\bibitem{DG}M. Drees and K. Grassie, Z.Phys. {\bf C28} (1985) 451.
\bibitem{H1N} T. Ahmed et al (H1 Collaboration), Zeit.f.Phys. {\bf C69} (1995)
27.
\bibitem{ZEUSN} M. Derrick et al (ZEUS Collaboration), Zeit.f.Phys. {\bf C63}
(1994) 391.
\bibitem{FS2} J. R. Forshaw and J. K. Storrow, Phys.Lett {\bf B321} (1994)
151.
\bibitem{H1} T.Ahmed et al (H1 Collaboration), Phys.Lett. {\bf B299}
(1993) 374.
\bibitem{ZEUS} M.Derrick et al (ZEUS Collaboration), Phys.Lett. {\bf B293}
(1992) 465.
\bibitem{data} D. O. Caldwell et al, Phys.Rev. {\bf D7}, 1362 (1973);
Phys.Rev.Lett. {\bf 40}, 1222 (1978).
\bibitem{Honjo} K. Honjo et al, Phys.Rev. {\bf D47}, (1993) 1048.
\bibitem{DL} A. Donnachie and P. V. Landshoff, Phys.Lett {\bf B296} 
227 (1992).
\bibitem{UA} C. Albajar et al (UA1 Collaboration), Nucl.Phys. {\bf B309}, 
405 (1988).
\bibitem{GG} R. Tanaka et al (AMY Collaboration), Phys.Lett. {\bf B277}, 215
(1992).
\bibitem{MRS}A. Martin, W.J. Stirling and R.G. Roberts,
Phys. Rev. {\bf D47} (1993) 867.
\bibitem{GRV}M. Gl\"uck, E. Reya and A. Vogt, Phys. Rev. {\bf D46}
(1992) 1973.
\bibitem{cone}J. Huth et al., Proc. of the 1990 DPF
Summer Study on High Energy Physics, Snowmass, Colerado,
ed. E.L. Berger (World Scientific, Singapore, 1992) p.134.
\bibitem{xgam} M. Derrick el al (ZEUS Collaboration),
Phys.Lett. {\bf B348} (1995) 665.
\bibitem{ZEUSbrus} Results from the ZEUS and H1 collaborations
presented by R. Brugnera in Proceedings of the International
Europhysics Conference on HEP, Brussels, 1995, submitted papers EPS-0380.
\bibitem{H1jets} S.Aid et al (H1 Collaboration), DESY 95-219.
\bibitem{steve}S. J. Maxfield, these proceedings.
\end{thebibliographyss}

\end{document}